\documentclass[10pt,preprint]{aastex}
\shorttitle{Spectroscopy of 2M0535B: Spots do not Cause Temperature Reversal}
\shortauthors{Mohanty \& Stassun}
\begin{document}
\def\teff{$T_{\rm eff}$}
\def\logg{$\log g$}
\def\zw{z_w }
\def\dop{\mathcal{D} }
\def\xangle{\vartheta }
\def\xanglew{\vartheta_w }
\def\Phid{{\Phi}_d }
\def\Phidx{{\Phi}_{\rm dx} }
\def\Phimx{{\Phi}_{\rm mx} }
\def\Phim{{\Phi}_{\rm m}}
\def\Phif{{\Phi}_{\rm f} }
\def\Phit{{\Phi}_t }
\def\Phiw{{\Phi}_w }
\def\Phir{{\Phi}_{\rm r} }
\def\rx{R_X }
\def\rk{R_k }
\def\del{{\bf{\nabla}} }
\def\delsq{{\nabla}^2 }
\def\vecr{{\bf{r}} }
\def\vecB{{\bf{B}} }
\def\vecBp{{\bf{B}_{\rm p}} }
\def\vecBa{{{B_{\varphi}}} }
\def\vecBr{{{B_{\varpi}}} }
\def\vecBz{{{B}_z} }
\def\malven{{\mathcal{M}}_A }
\def\rstar{R_{\ast} }
\def\rd{R_D }
\def\omstar{{\Omega}_{\ast} }
\def\omx{{\Omega}_X }
\def\ax{a_X }
\def\jbstar{\bar J_{\ast} }
\def\zmax{z_{\rm max} }
\def\deg{$^{\circ}$ }
\def\erhat{{\^{e}}$_{\varpi}$}
\def\eahat{{\^{e}}$_{\varphi}$}
\def\ezhat{{\^{e}}$_z$}
\def\mstar{M_{\ast} }
\def\rstar{R_{\ast} }
\def\lstar{L_{\ast} }
\def\msun{M_{\odot} }
\def\rsun{R_{\odot} }
\def\lsun{L_{\odot} }
\def\mdotd{\dot M_D }
\def\mdot{\dot M }
\def\rhos{{\rho}_0 }
\def\vs{v_0 }
\def\rhoa{{\rho}_{Am} }
\def\va{v_{Am} }
\def\bs{B_0 }
\def\ba{B_{Am} }
\def\rs{r_0 }
\def\ra{r_{Am} }
\def\rr{r_{A2} }
\def\vp{{\mathcal{V}}_0 }
\def\h2{{\rm H}_2 }
\def\met{{\rm M}^{+} }
\def\avz{\langle Z \rangle }
\def\gz{{\rm gr}\avz }
\def\gn{{\rm gr}0 }
\def\gneg{{\rm gr}^{-} }
\def\nh2{n_{\h2} }
\def\ne{n_e }
\def\ni{n_i }
\def\ngz{n_{\gz} }
\def\ngn{n_{\gn} }
\def\ngneg{n_{\gneg} }
\def\xe{x_e }
\def\xi{x_i }
\def\xgz{x_{\gz} }
\def\xgneg{x_{\gneg} }
\def\aie{\alpha_{i,e} }
\def\agze{\alpha_{e,\gz} }
\def\agzi{\alpha_{i,\gz} }
\def\agne{\alpha_{e,\gn} }
\def\agni{\alpha_{i,\gn} }
\def\agnq{\alpha_{q,\gn} }
\def\zetax{\zeta_X }
\def\col{N_{\perp} }
\def\baraie{{\bar{\alpha}}_{i,e} }
\def\xgcross{x_{\rm gr}^{\ast} }

%\title{The Young Substellar Eclipsing Binary 2MASS 0535-0546. II. \\ No Evidence that Spots Cause the Temperature Reversal}
\title{High-Resolution Spectroscopy During Eclipse of the Young
Substellar Eclipsing Binary 2MASS~0535$-$0546.\ II.\ Secondary Spectrum:
No Evidence that Spots Cause the Temperature Reversal}
\author{Subhanjoy Mohanty\altaffilmark{1}, Keivan G.\ Stassun\altaffilmark{2,3}}
\altaffiltext{1}{Imperial College London, 1010 Blackett Lab., Prince Consort Road, London SW7 2AZ, UK.  {\tt s.mohanty@imperial.ac.uk}}  
\altaffiltext{2}{Department of Physics \& Astronomy, Vanderbilt University, Nashville, TN 37235, USA.  {\tt keivan.stassun@vanderbilt.edu}}  
\altaffiltext{3}{Department of Physics, Fisk University, Nashville, TN 37208, USA.}  

\begin{abstract}
We present high-resolution optical spectra of the young
brown-dwarf eclipsing binary 2M0535$-$05, obtained during
eclipse of the higher-mass (primary) brown dwarf.  Combined with our
previous spectrum of the primary alone (Paper~I), the new observations yield the spectrum of the 
secondary alone.  We investigate, through a differential
analysis of the two binary components, whether cool surface spots are
responsible for suppressing the temperature of the primary.  In Paper~I, we found a significant
discrepancy between the empirical surface gravity of the primary
%(derived from the orbital dynamics and lightcurves), 
and that inferred via fine analysis of its spectrum.
%the TiO-$\epsilon$ band and the red lobe of the \ion{K}{1} doublet.  
%We ascribed this to either
%a large surface coverage by cool spots, or opacity uncertainties
%in the synthetic models used in the spectral analysis.  
%We now
%perform exactly the same fine analysis with the secondary's spectrum.
Here we find precisely the {\it same} discrepancy in surface gravity, both
qualitatively and quantitatively. While this
may again be ascribed to either cool spots or model opacity errors, it implies
that cool spots {\it cannot} be responsible for {\it preferentially}
lowering the temperature of the primary: if they were, spot effects on
the primary spectrum should be preferentially larger, and they are not.
The \teff\ we infer for the primary and secondary, from the
TiO-$\epsilon$ bands alone, show the same reversal, in the same ratio,
as is empirically observed, bolstering the validity of our analysis.
In turn, this implies that if suppression of convection by magnetic
fields on the primary is the fundamental cause of the \teff\ reversal,
then it cannot be a local suppression
yielding spots mainly on the primary (though both components may be
equally spotted), but a {\it global} suppression in the interior of the
primary.  We briefly discuss current theories of how this might work.
\end{abstract}


\section{Introduction: Results for 2M0535 So Far}
2MASS 05352184$-$0546085 (hereafter 2M0535) is a very young system
located in the Orion Nebula Cluster, and identified by \citet[][hereafter SMV06]{stassun06} 
as the first known substellar eclipsing
binary (EB).  EBs allow extremely precise direct measurements (via
their orbital dynamics and eclipse lightcurves) of the component masses
and radii, and hence their surface gravities ($\log g$), as well as the
ratio of their luminosities (and thus the ratio of their effective
temperatures, \teff).  2M0535 therefore enables one to rigorously
test the evolutionary models and synthetic spectra that are widely
used to characterize the vast majority of brown dwarfs (for which a
direct determination of mass and radius is not possible), and to do
so at the young ages where our theoretical knowledge is most lacking.

The measurements by SMV06 were refined by \citet{stassun07}
and \citet[][hereafter G09]{gomez09}. 
The central results are:
{\it (1)} Both components of 2M0535 are moderate mass brown dwarfs
($M_{\rm A} = 0.0572 \pm 0.0033 \,\msun$, $M_{\rm B} = 0.0366 \pm
0.0022 \,\msun$); {\it (2)} their radii ($R_{\rm A} = 0.690 \pm
0.011 \rsun$, $R_{\rm B} = 0.540 \pm 0.009 \rsun$) are consistent
with the theoretical expectation that young brown dwarfs 
should be much larger than their field counterparts; and {\it
(3)} the \teff\ ratio of the components ($T_{\rm eff,B}/T_{\rm eff,A}
= 1.050 \pm 0.004$) shows a surprising reversal, with the more massive
primary (A) being cooler than the secondary (B).

The \teff\ reversal is not predicted by any current set of theoretical
evolutionary tracks.  To explain it, \citet[][henceforth CGB07]{chabrier07} 
proposed that strong magnetic fields on the primary
suppress convection, both globally in the interior, and/or locally near
the surface (producing cool surface spots); neither effect is included
in standard evolutionary models, and both would act to depress the
effective temperature of the primary.  \citet[][hereafter MM09]{macdonald09} 
subsequently put forward a qualitatively similar
hypothesis, wherein magnetic fields lower the \teff\ of the primary
by inhibiting interior convection (though their theory differs in
important respects from that of CGB07, as we discuss later).

Bolstering the case for strong magnetic fields preferentially affecting
the primary, \citet{reiners07} found that, compared to the
secondary, the primary is a relatively rapid rotator (which should
boost field generation), with very prominent chromospheric H$\alpha$
emission (ultimately powered by the release of magnetic stresses).
\citet{mohanty09} subsequently showed, through an analysis of
the optical to mid-infrared spectral energy distribution of 2M0535,
that ongoing disk accretion is highly improbable in this system.
Thus the H$\alpha$ emission in the primary is indeed likely to be
chromospheric in origin, supporting Reiners et al.'s conclusion that
it harbors strong magnetic fields.  Interestingly, these results are
also consistent with the behaviour of low-mass {\it field} stars:
chromospherically active late-K and M field dwarfs appear cooler than
inactive ones of the same luminosity \citep{morales08,stassun12}.

The implication is that magnetic fields appear to have a large
impact on the effective temperature of low-mass stars and brown
dwarfs.  If true, this would have far-reaching consequences for our
understanding of these objects.  For example, the initial mass function
(IMF) in young star-forming regions is often derived by comparing
(sub)stellar luminosities and temperatures to theoretical HR diagrams.
The latter do not include any field effects, while a substantial
fraction of low-mass stars and brown dwarfs at these very early ages
show evidence of strong fields (rapid rotation and high activity),
just like 2M0535A and active field dwarfs; thus, the inferred IMF
might be skewed by a field-induced depression in \teff, potentially
causing an overestimate of the number of low-mass brown dwarfs.

The issue now is to decipher how exactly fields achieve this
effect, if indeed they are responsible.  As noted above, theory
suggests they might do so by producing cool surface spots (through
local suppression of convection), and/or by inhibiting convection
globally in the (sub)stellar interior.  Observationally, checking
for the presence of spots is clearly easier.  G09
carried out the first investigation of spots on
2M0535AB.  They showed that the small amplitude residual (non-eclipse)
variations in the system's lightcurve, modulated at the rotational
periods of the primary and secondary, can be well-reproduced by cool
spots asymmetrically covering a small fraction ($\lesssim$ 10\%)
of both components' surfaces.  While they find no evidence for a
large ($\gtrsim$ 50\%) spot coverage preferentially on the primary,
as required by CGB07's theory if spots are to account for the \teff\
reversal, they cannot rule out such spots either, if the latter are
arranged {\it symmetrically} about the primary's rotation axis (e.g.,
polar spots, latitudinal bands, or ``leopard spots'').

To check for spots independent of their orientation, one must examine
the spectra of the binary components.  Cool spots are by definition
cooler than the surrounding photosphere, and the effective surface
gravity within them is also lower (because the magnetic pressure partly
offsets the gas pressure, mimicking the reduction in gas pressure
caused by a lower surface gravity); both effects alter the shape of
temperature- and gravity- (more accurately, gas-pressure-) sensitive
spectral lines relative to an unspotted spectrum.  We embarked upon
this study in a previous paper \citep[][hereafter Paper~I]{mohanty10}, 
where we analysed the high-resolution optical spectrum of
the primary 2M0535A alone (obtained during secondary eclipse, with
a negligible 1.6\% contamination by the secondary). Specifically,
we derived \teff\ and $\log g$ by simultaneously fitting the
observed TiO-$\epsilon$ band and red lobe of the KI doublet with
state-of-the-art synthetic spectra, and compared our results to the
empirically known $\log g$.  We found that at the \teff\ $\approx$
2500 K inferred from TiO, the KI lobe implies $\log g$ $\approx$ 3.0,
or 0.5 dex lower than the empirical value.  Conversely, at the known
$\log g$ of 3.5, the \teff\ inferred from \ion{K}{1} is 2650 K, or 150 K higher
than derived from TiO.

Such discrepancies are indeed expected if the photosphere is spotted,
due to the temperature and effective gravity properties of spots
noted above.  In particular, we showed that the spectrum of 2M0535A
is consistent with an unspotted stellar photosphere with \teff\ =
2700 K and (empirical) $\log g$ = 3.5, coupled with axisymmetric cool
spots that are 15\% cooler (2300 K), have an effective $\log g$ =
3.0 (0.5 dex lower than photospheric) and cover 70\% of the surface.
The spot temperature and gravity are consistent with the properties
of sunspots and starspots in general, as well as with the previous
lightcurve analysis of 2M0535, while the covering fraction agrees
with CGB07's requirement if spots are to cause the \teff\ reversal.

On the other hand, these discrepancies may arise from errors
in the molecular opacities or equation of state (EOS) in the
synthetic spectra we used to fit the data.  Such errors are known to be
present from analyses of field dwarfs \citep{reiners05};
while it is unclear whether they persist in the model spectra
at the much lower \logg\ appropriate for very young brown dwarfs,
it is certainly possible (see Paper~I).

\section{Goal of This Paper\label{goals}}
In either case, a resolution demands an analogous analysis of the
spectrum of the secondary, 2M0535B.  That is our goal here.  If a
large covering fraction of spots is preferentially affecting the
primary, then the much less spotted secondary should evince much
smaller spectral discrepancies, if any.  If opacity/EOS errors are to
blame instead, then the secondary should show similar discrepancies:
since its empirical $\log g$ is nearly the same as the primary's,
and so is its \teff\ (as implied by an empirical \teff\ ratio very
close to unity), opacity/EOS uncertainties in the model spectra
should affect our analyses of both components to a similar degree.
Note that, in the latter case, one cannot appeal to real spot
effects on the primary simply being washed out by model errors.
As discussed above, the primary spot covering fraction {\it required}
to produce the \teff\ reversal is of the same order as inferred from
its spectrum under the assumption of error-{\it free} models; if model
errors are in fact large enough to overwhelm spot effects, then such
spots are automatically too small to cause the reversal.  Finally,
the secondary may also exhibit similar (or larger) discrepancies if
it is as (or more) spotted than the primary.  In this case too, spots
can be excluded as the {\it cause} of the \teff\ reversal, since the
latter requires the primary to be far more spotted than the secondary.

To summarize, our aim is to investigate whether the secondary spectrum
shows much smaller discrepancies than the primary; if it does, then
spots on the primary are favoured as the mechanism for \teff\ reversal,
otherwise such spots can be effectively ruled out.

\section{Observations and Data Reduction}
The data collection and reduction for the observation of the 
primary eclipse were performed in precisely the same manner, using
the same instrumental setup and procedures, as for our previous
observation and analysis of the secondary eclipse (Paper~I). Here we
briefly summarize the salient details.

We observed 2M0535 on the night of UT 2011 March 15 with the High
Resolution Echelle Spectrometer (HIRES) on Keck-I\footnote{Time
allocation through NOAO via the NSF's Telescope System Instrumentation
Program (TSIP).}.  We observed in the spectrograph's ``red"
(HIRESr) configuration with an echelle angle of $-0.403$~deg and
a cross-disperser angle of 1.7035~deg. In this configuration,
the two features of primary interest in this paper, TiO
$\lambda\lambda$8435--8455 and \ion{K}{1} $\lambda$7700, fall on the
``green" chip, in echelle orders 42 and 46, respectively.  We used the
OG530 order-blocking filter and the 1\farcs15$\times$7\farcs0 slit, and
binned the chip during readout by 2 pixels in the dispersion direction.
The resulting resolving power is $R\approx 34\, 000$, with a 3.7-pixel
($\sim 8.8$ km s$^{-1}$) FWHM resolution element.

We obtained three consecutive integrations of 2M0535, each of 2400~s.
ThAr arc lamp calibration exposures were obtained before and after
the 2M0535 exposures, and sequences of bias and dome flat-field exposures
were obtained at the beginning of the night. The 2M0535 exposures were
processed along with these calibrations using standard
IRAF\footnote{IRAF is distributed by the National Optical Astronomy
Observatory, which is operated by the Association of Universities for
Research in Astronomy (AURA) under cooperative agreement with the National
Science Foundation.}
tasks and the {\sc makee} reduction package written for HIRES by
T.~Barlow. The latter includes optimal extraction of the orders as well
as subtraction of the adjacent sky background.  The three exposures
of 2M0535 were processed separately and then median combined with
cosmic-ray rejection into a single final spectrum.  The signal-to-noise
(S/N) of the final spectrum is $\sim 15$ per resolution element.

To complement our observation of the secondary eclipse in Paper~I,
here we intentionally chose the observations to coincide
exactly with the primary eclipse, i.e.\ when the higher-mass,
larger, lower-\teff\ primary component was behind the secondary
as seen from Earth. The first exposure started at UT 05:51~hr, and
the third exposure ended at UT 07:54~hr, corresponding to orbital
phases of 0.7371 and 0.7458, respectively, during which time the
primary is maximally blocked \citep[cf.\ Fig.\ 3 in][]{stassun07}.
Integrated over the entire 2-hr observation,
the total light contribution from the blocked primary was $\approx$35.1\%,
calculated using the
accurately determined radius ratio, temperature ratio, and orbital
parameters, including the orbital inclination, from the light curve
modeling performed in G09.
 
Given the exorbitant time cost of the observations and the difficulty
in scheduling at precisely the right phase, it
was not feasible to obtain multiple observations for this project. However,
the extensive light curve observations of \citet{stassun06,stassun07}
and G09, spanning more than 10 years, clearly
demonstrate that the system is not variable outside of eclipse at more
than $\sim$10\%. So we have good reason to believe that the observation 
presented here should be adequately representative.

\section{Methodology}

\subsection{Isolation of the secondary spectrum\label{isolation}}
For our analysis, we require the spectrum of the lower-mass component
(the ``secondary") of 2M0535 alone, i.e., as uncontaminated by light 
from the primary component as possible. In Paper~I, we observed the
2M0535 system during secondary eclipse, i.e., when the primary 
component was in front of the secondary. The secondary eclipse
is near total, and thus our observations in Paper~I provided a ``pure"
measurement of the primary component's spectrum, with a negligible 1.6\%
contamination from the secondary component.

The spectrum that we have obtained here of the secondary component is
much less pure---it is 35.1\% contaminated by the primary's light (see
above)---because the primary eclipse is not total. Thus, we have used
our isolated spectrum of the primary component from Paper~I to correct
the observed spectrum of the secondary component. Specifically, we
shifted the isolated primary spectrum in velocity according to the known
ephemeris of the system (see G09), scaled it to 35.1\% of the total 
light, and subtracted it from the observed spectrum of the secondary.
Because this procedure involves subtraction of two observed spectra
with comparable S/N, the S/N of the resulting final isolated spectrum 
of the secondary is decreased to $\sim 10$.

\subsection{Synthetic spectra} 
In order to conduct an analysis identical to that in Paper~I, we use
a subset of the same atmospheric models adopted in that paper.
Specifically, we use synthetic spectra for plane-parallel atmospheres 
generated using the PHOENIX code, designated AMES-Cond 
\citep[version 2.4][]{allard01}. 
As in our analysis of the primary, we use solar-metallicity 
models ($[M/H] = 0.0$). While the metallicity of 2M0535 is not 
explicitly known, a large deviation from solar is not expected for a 
young object in a nearby star-forming region. 

Note that in Paper~I, we considered AMES-Dusty as well as AMES-Cond models. Both treat the formation of dust grains self-consistently, through chemical equilibrium calculations.  Once formed, however, the grains are assumed to remain entirely suspended in the photosphere in the Dusty models, and settle completely below the photosphere in the Cond ones.  Under physical conditions where the chemical equations imply {\it no} dust formation, the Cond and Dusty spectra are identical; in the models, this occurs for \teff\ $\ge$ 2500~K.  For the latter temperatures, therefore, either set of models may be used.  In Paper~I, we used Dusty models for \teff\ $<$ 2500~K (i.e., in the dust formation regime; for late M-types with grains, Dusty models are expected to be more appropriate than Cond, since grains are expected to start settling out of the atmosphere only around mid-to-late L types), and Cond models for \teff\ $\ge$ 2500~K (since either Cond or Dusty can be used in this grainless regime, and the high-resolution Cond spectra available to us extend to higher \teff\ than the Dusty ones).  In the present paper, concerned with investigating the secondary component of 2M0535, we use only the (grainless) Cond models at \teff\ $>$ 2500~K, since the secondary is {\it warmer} than the primary considered in Paper~I, and is definitely hotter than $\sim$2500~K \citep[see][]{mohanty09}.

\subsection{Determination of \teff\ and \logg}
Our goal is to determine the \teff\ and \logg\ of the lower-mass
component of 2M0535 (the ``secondary," hereafter 2M0535B) from comparisons
to synthetic spectra. 
As in Paper~I, we again follow
\citep[][hereafter M04]{mohanty04}, who have
shown that two ideal regions for this analysis are the TiO- bandheads at
$\lambda\lambda\lambda$8435, 8445, 8455, and the red lobe of the 
\ion{K}{1}doublet at $\lambda$7700
(the blue lobe falls in the gap between echelle orders in the HIRES
setting used). In particular, the TiO bandheads are very sensitive
to \teff, but negligibly so to \logg, while the \ion{K}{1} absorption is
sensitive to both; using the two regions in tandem therefore enables
one to disentangle and individually determine these two parameters, 
as we did for the 2M0535A component in Paper~I.

The synthetic spectra were treated for comparison
to the data precisely as in Paper~I.  Briefly, the model spectra were broadened by the instrumental profile, and interpolated onto the wavelength scale of the data.  Since our data are not flux calibrated, both the data and models were then normalized over a selected pseudo-continuum interval, just outside the TiO-$\epsilon$ and \ion{K}{1} bands of interest, for comparison: over $\lambda\lambda$[8405.0--8414.0] for TiO, and $\lambda\lambda$[7693.0--7698.0] for \ion{K}{1}.  The only change from Paper~I is that no rotational broadening of the synthetic spectra was required in the present case, since the slowly-rotating 2M0535B is effectively unbroadened 
\citep[$v \sin i < 5$ km~s$^{-1}$;][]{reiners07}
compared to the instrumental resolution (8.8 km~s$^{-1}$; see above).  
Finally, we note, as in Paper~I, that the models were originally
constructed at intervals of 100~K and 0.5~dex in \teff\ and \logg,
respectively, so we have linearly interpolated between adjacent spectra
to construct a finer final grid of models,
with steps of 50~K in \teff\ and 0.25~dex in \logg.

\section{Results\label{results}}
\begin{figure}
\includegraphics[scale=0.7]{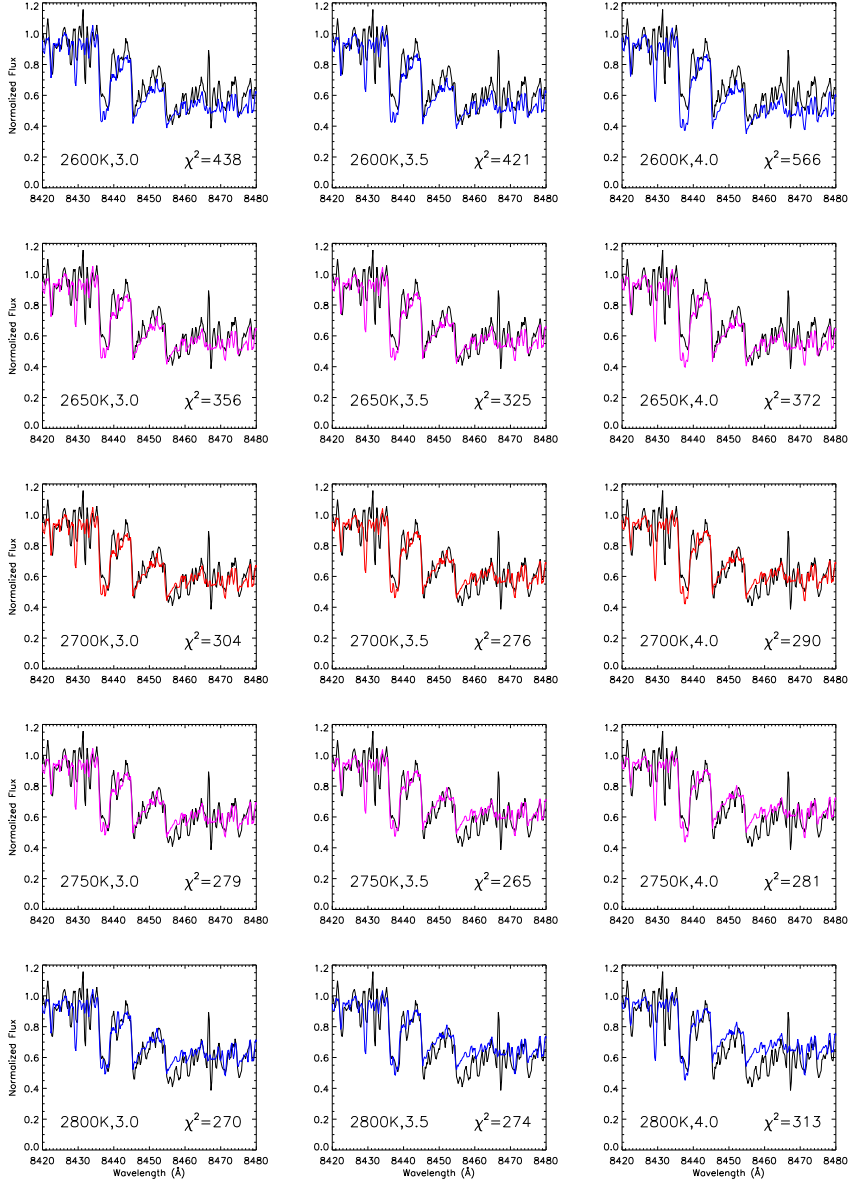}
\caption{\label{fig:tiogrid}
Observed TiO-$\epsilon$ region in 2M0535B (black) compared to Cond models \citep{allard01}. 
Best-fit models shown in red; worse but still admissible fits by eye shown in magenta; and all others, 
which clearly diverge from the data by eye, shown in blue. 
The by-eye fits prefer \teff\ $\approx$ 2700$\pm$50 K.
Note that the model fits are relatively insensitive to gravity over the 1 dex range plotted; see Section~\ref{results}.  The corresponding $\chi^2$ values for the model fits are also shown in each panel.  Note that the best-fit $\chi^2$ models (2700--2800\,K) are $\sim$50\,K hotter than the best-fit models by eye (2650--2750\,K); this is because the cores of the bandheads are better fit by the slightly cooler models, but the flux in these spectral regions is weighted less in the $\chi^2$ fits to account for lower S/N there.  The best-fit models from the two methods are nevertheless consistent with each other, given the $\sim$50\,K uncertainty in the by-eye fits; see \S5.   }
\end{figure}

\begin{figure}
\includegraphics[scale=0.5]{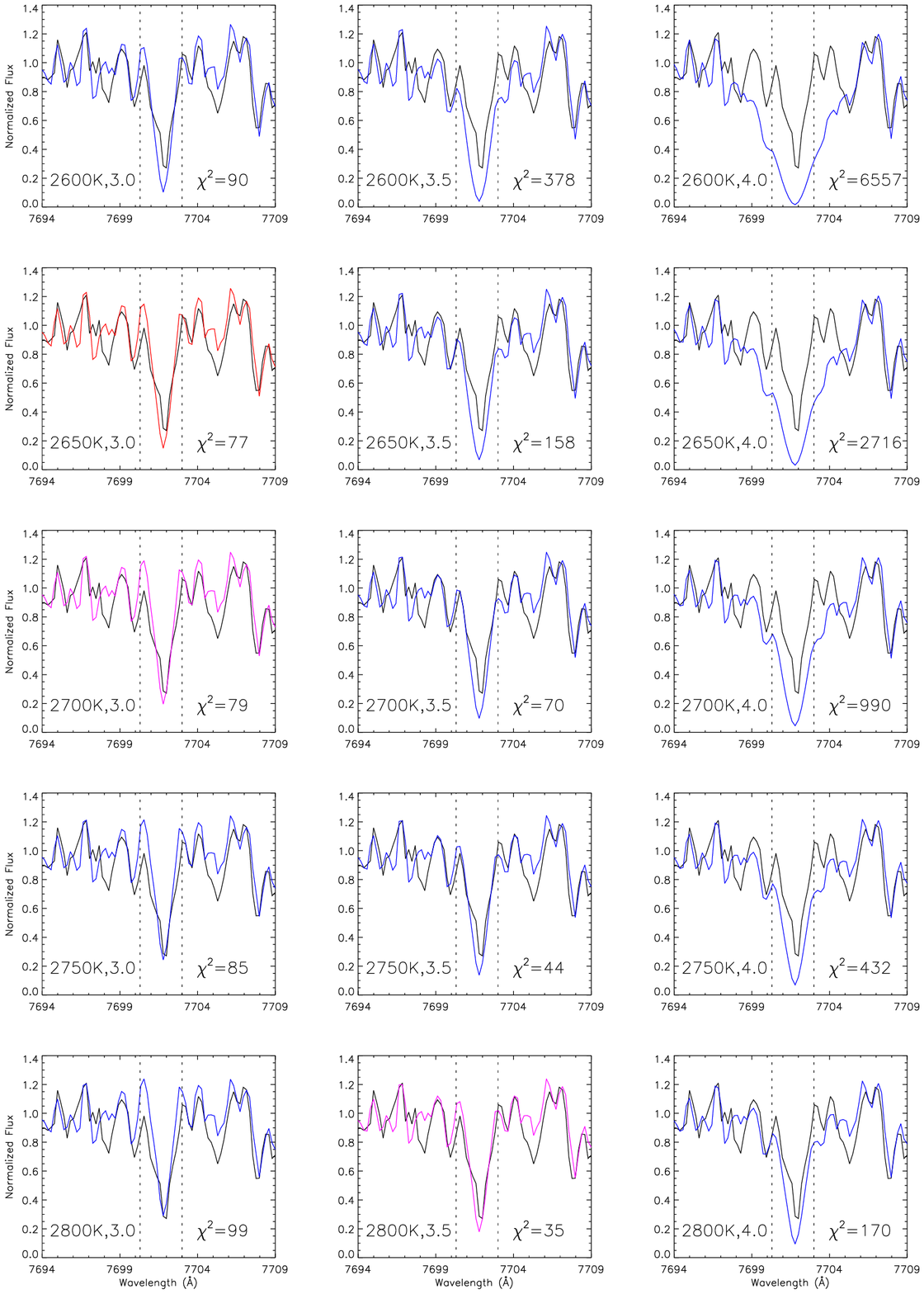}
\includegraphics[scale=0.5]{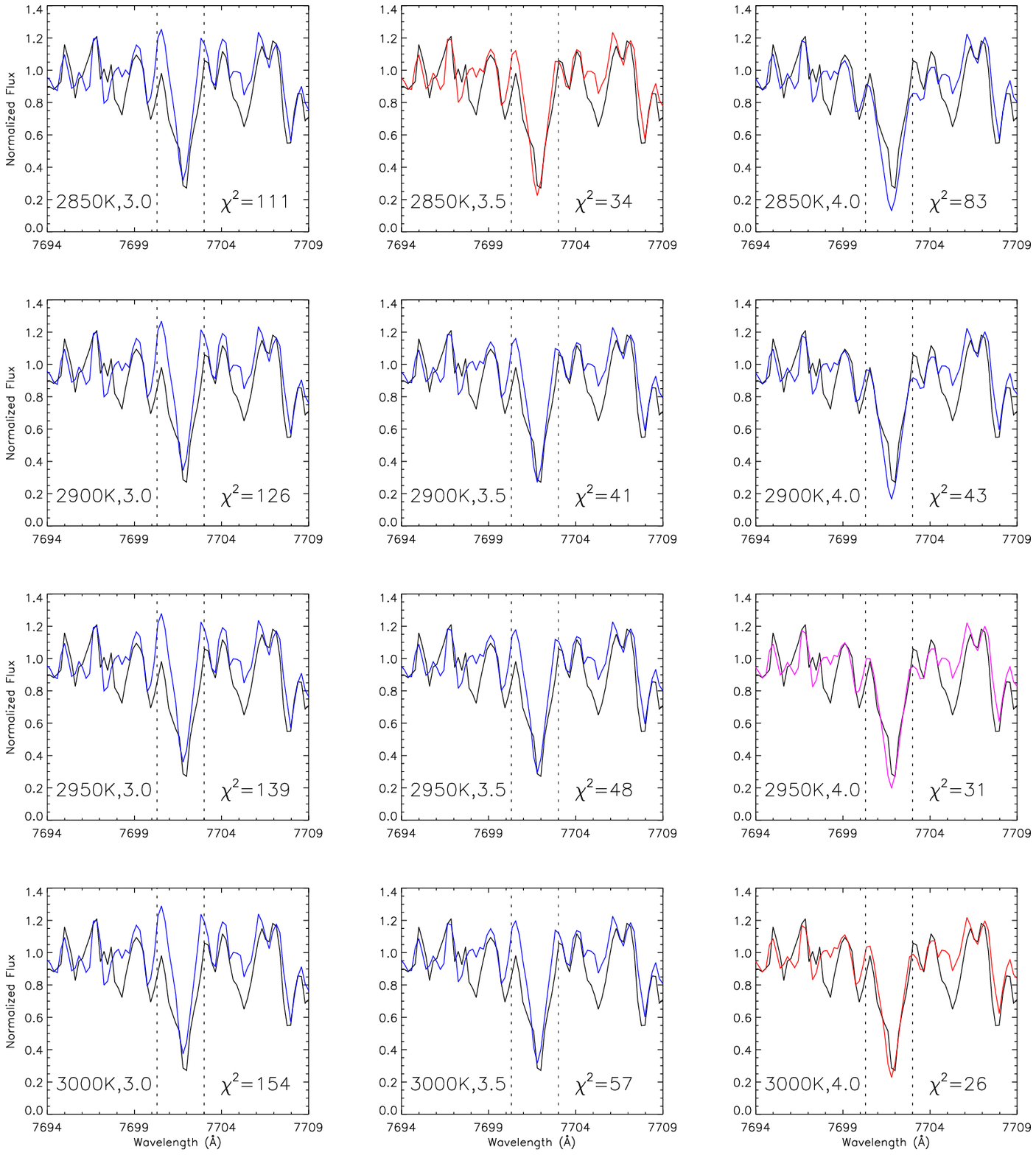}
\caption{\label{fig:kigrid}
Observed red lobe of \ion{K}{1} in 2M0535B (black) compared to Cond models \citep{allard01}. 
From top to bottom, the three columns on the left depict model \teff\ from 2600 to 2800\,K, and the three columns on the right show \teff\ from 2850 to 3000\,K.  
Best-fit models by eye are shown in red; 
worse but still admissible fits by eye are shown in magenta; and all others, which clearly diverge from the data, 
are shown in blue. 
Note that the \ion{K}{1} absorption is sensitive to both \teff\ and gravity: 
a 150 K increase in \teff\ compensates for a 0.5 dex rise in log g. 
At \teff\ = 2700 K, corresponding to the best-fit by eye to TiO-$\epsilon$ (Figure \ref{fig:tiogrid}), 
the \ion{K}{1} by-eye fit implies $\log g$ = 3.0, while at the empirically determined $\log g$ = 3.5, it implies 
\teff\ = 2850 K.  At the still higher gravity of \logg\ = 4.0, the best model fit by eye is at 3000\,K, but this gravity is empirically ruled out. The corresponding $\chi^2$ values are also shown in each panel; the best-fit models implied by these values are in good agreement with the by-eye results.  See Section~\ref{results}.}
\end{figure}

The models are overplotted on the data in Figs.\,\ref{fig:tiogrid} and \ref{fig:kigrid}, for TiO and \ion{K}{1} respectively, over the range of model \teff\ and \logg\ plausible for this cool and very young (low-gravity) object.  To guide the eye better, both the synthetic and observed spectra have been boxcar-smoothed by 3 pixels in these plots.  Models that best fit the data, by eye, are plotted in red, marginally worse fits in magenta, and bad fits in blue.  We note the following trends.

\noindent {\it {TiO-$\epsilon$}}:  Fig.\,\ref{fig:tiogrid} shows that, as expected, the TiO-$\epsilon$ bandheads are quite insensitive to gravity, over the range in \logg\ shown, but highly sensitive to temperature.  The best fit by eye is obtained to the model at 2700\,K (in red), while models within $\pm$50\,K of this (in magenta) are marginally worse.  Models that are even cooler/hotter (in blue) appear clearly inconsistent with the data: the bandheads at 8445\,\AA\ and (especially) 8435\,\AA\ are significantly stronger than the data by \teff\ = 2600\,K, while the bandheads at 8445\,\AA\ and (especially) 8455\,\AA\ are considerably weaker than the data by 2800\,K.  Since even a 100\,K deviation from the best fit is evident to the eye, our precision in \teff\ determination by eye is likely to be $\sim$50\,K (in agreement with M04).  From the TiO-$\epsilon$ fits alone, therefore, we would infer \teff\ $\approx$ 2700$\pm$50\,K.      

\noindent {\it {\ion{K}{1} and TiO-$\epsilon$}}:  Fig.\,\ref{fig:kigrid} reveals, again as expected, that the \ion{K}{1} line is very sensitive to both \teff\ and \logg, becoming stronger with both decreasing temperature and increasing gravity (the extent of this absorption line -- [7700--7703]\,\AA\,, used for both these by-eye fits and the chi-squared analysis below -- is demarcated by the vertical dashed lines in this plot).  We see that, at the temperature inferred from TiO (2700$\pm$50\,K), the best-fit by eye to \ion{K}{1} is at \logg\ = 3.0 (the best fit here is actually at the lower end of this \teff\ range, at 2650\,K (red); the 2700\,K model \ion{K}{1} (magenta) is somewhat narrower than the data, while the 2750\,K model (blue) is appreciably narrower).  At the empirically known gravity of this object, \logg\ = 3.5, 2700$\pm$50\,K models are significantly stronger (broader and deeper) than the data; at this gravity, the best by-eye model fit is obtained at 2850\,K (red), which is unsupported by the TiO fits.  Finally, for completeness, we also show that at \logg\ = 4.0, the best fit by eye is at 3000\,K (red).  The steadily increasing best-fit \teff\ with increasing \logg\ simply illustrates the degeneracy between temperature and gravity in the \ion{K}{1} line: specifically, an increase of 0.5 dex in \logg\ is compensated for by an increase of $\sim$150--200\,K in \teff\ (as found my M04).  In summary, the best fit to \ion{K}{1} at the \teff\ derived from TiO occurs at a gravity 0.5\,dex lower than the known value, while imposing the correct \logg\ yields a \teff\ that is $\sim$150\,K higher than from TiO, and incompatible with the latter.  %As an aside, we note that the [2800\,K, 3.5] model appears to be a superior fit than those at any other values.           

To better quantify the fitting results from Figs.\,\ref{fig:tiogrid} and \ref{fig:kigrid}, and the uncertainties therein, we carried out chi-square comparisons between the synthetic spectra and data.  The TiO comparisons were made over the wavelength range [8420:8480]\,\AA\ (which includes all three bandheads; see Fig.\,\ref{fig:tiogrid}), and the \ion{K}{1} ones over [7700.2:7703]\,\AA (corresponding to the entire line, until the pseudo-continuum is reached on either side of line center; see Fig.\,\ref{fig:kigrid}).  These ranges correspond to 26 data points for \ion{K}{1} and 265 for TiO.  The data and models were {\it not} smoothed for this exercise, since smoothing introduces correlations between adjacent pixels, thereby vitiating the interpretation of the chi-square vaues.  However, the S/N of our data for 2M0535B is lower than that for the primary in Paper~I (mainly because, while we compensated for the lower luminosity of the secondary with longer observing times, isolating the secondary spectrum required subtracting the scaled primary spectrum from our data, adding to the noise; see \S\ref{isolation}).  This is of concern in the line/bandhead cores, where the flux is lowest, and especially in the \ion{K}{1} line core, which is significantly fainter, and thus noisier, than the bottom of the TiO bandheads.  To account for this, our chi-square values for individual pixels are weighted by the data pixel flux (for both TiO and \ion{K}{1}), effectively making fainter pixels (and thus the line cores) less significant than brighter ones.

\begin{figure}
\includegraphics[scale=0.7]{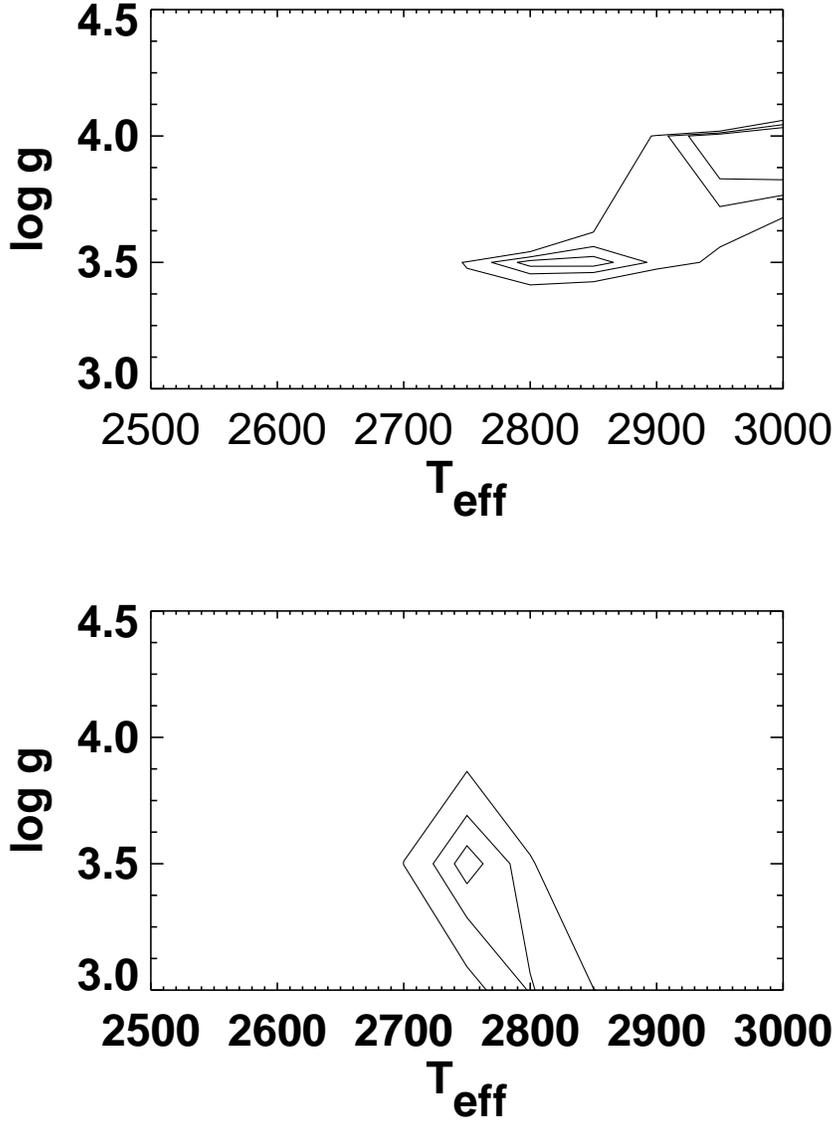}
\caption{\label{fig:chisq}
Determination of goodness-of-fit and formal fit parameter
uncertainties. Contours of constant 
$\chi^2 - \chi_{\rm min}^2 =$ 2.3, 6.2, 11.8,
representing 1$\sigma$, 2$\sigma$, 3$\sigma$ joint confidence 
intervals in the \teff--\logg\
parameter plane. Top: joint confidence intervals for fitting of \ion{K}{1}.
The absolute minimum $\chi^2$ best fit is for \teff\ = 3000~K and \logg\ = 4.0, however an equally good 
fit at the 1$\sigma$ level is at \teff\ = 2825~K and \logg\ = 3.5 (see Fig.~\ref{fig:kigrid}). 
%however a second equally good fit within 1σ confidence
%occurs at Teff = 2650 K and log g = 3.5. 
Bottom: joint confidence intervals for fitting of TiO. 
%The contours demonstrate that for TiO
%the best-fitting model spectra are relatively insensitive to log g
%but highly sensitive to Teff; 
A best-fit \teff\ = $2750 \pm 15$ K is strongly preferred at high confidence.
The best-fit \teff\ from TiO-$\epsilon$ and \ion{K}{1} at the empirical \logg\ of 3.5 differ
by 100--150~K at $\sim 3\sigma$ confidence.}
\end{figure}

The results are plotted in Fig.\,\ref{fig:chisq}.  At the empirical gravity of 2M0535, (\logg\ = 3.5), the best-fit \teff\ from \ion{K}{1} is 2825~K, in agreement with our by-eye estimate above\footnote{Note that the other peak at [$\sim$3000\,K, 4.0] is formally the lowest $\chi^2$ in our fits, but consistent at 1$\sigma$ with the fit at [2825~K, 3.5]; this simply illustrates that \ion{K}{1} is roughly degenerate in \teff\ and \logg\ (increasing \teff\ balances increasing \logg), as noted above and discussed in detail by M04.  Indeed, there may be equally good fits to \ion{K}{1} at even higher \teff\ and \logg, due to this degeneracy; we have not explored this parameter space because such high \logg\ are empirically ruled out for the secondary.}. The corresponding 1$\sigma$ uncertainty is 35~K.  The TiO chi-squares indicate \teff\ = $2750 \pm 15$~K (also at the empirical \logg\ = 3.5, though these chi-square contours discriminate much less between different gravities than the \ion{K}{1} contours, in keeping with the relative insensitivity of TiO to \logg).  This is again consistent with the by-eye \teff\ inferred from TiO\footnote{Note that the best chi-square fits to TiO (see chi-square numbers stated in the Fig.\,1 panels) imply temperatures 50\,K higher than implied by the best-fits by eye (shown by the red and purple models in Fig.\,1).  This is because the eye is drawn to (dis)agreement between models and data in the core of the bandheads, while in the chi-square fits, we have explicitly given lower weight to the fluxes here to account for lower S/N, as discussed above.  Nevertheless, the best chi-square fit remains consistent with the best-fits by eye, within the $\sim$50\,K errors in the latter.}.  Finally, at this \teff\ and \logg\ = 3.5, the \ion{K}{1} line is marginally fit at the 3$\sigma$ contour level.  Given the formal uncertainties in \teff\ from the TiO and \ion{K}{1} chi-squares, the probability that the two \teff\ at this gravity are equal is only 0.7\%.  We can thus rule out the possibility that TiO and \ion{K}{1} give the same \teff\ (at the same, empirically determined, \logg) with 99.3\% confidence.  
 
These results are essentially the same as those obtained in Paper~I for 2M0535A: in both the primary and the secondary, the temperatures obtained from TiO-$\epsilon$ and \ion{K}{1} are incompatible with each other by $\sim$100--150\,K, which is statistically significant at the $\geq$3$\sigma$ level.    

We caution that the statistical significance of this discrepancy is formally higher for the primary (see Paper~I), because of the lower S/N (equivalently, lower chi-square weighting of the line cores) in the secondary.  Thus, our results here should be validated further with higher S/N observations of 2M0535B.  Our confidence that our results are not due simply to statistical fluctuations, however, is bolstered by three facts: {\it (a)} the qualitative trend for both components is the same: the \teff\ from TiO is lower than from \ion{K}{1}; {\it (b)} the quantitative discrepancy between the TiO and \ion{K}{1} temperatures is also roughly the same in both cases; and {\it (c)} the ratio of component \teff\ we obtain from TiO, \teff$_B$/\teff$_A$ $\approx$ 1.1, is consistent (within the $\pm$50\,K spacing in our model \teff\ grid) with the empirical ratio of 1.05.    

%We note here that, while the best-fit chi-square values are the same
%as obtained by eye, the formal uncertainties cited above for the
%chi-square analysis, obtained via interpolation over the model grid,
%are smaller than the model grid spacing (50~K in \teff\
%and 0.25 dex in \logg). We therefore adopt the grid spacing of 50~K
%and 0.25 dex as a conservative estimate of our uncertainties for the
%rest of the paper (corresponding to the same errors assumed for the
%fits by eye).

\section{Discussion}

\subsection{Implications for Spots Causing \teff\ Reversal}
Assuming that our results are not substantially affected by the lack of high S/N data in the very cores of the lines in the secondary spectrum (as we have argued above they are not), there are only four possible interpretations of the combined analysis presented here and in Paper~I. 

\noindent {\it (1)}\,  Spots are the major cause of the spectral
discrepancy in 2M0535A and B\footnote{As an aside, we note that this
scenario is unlikely: {\it Both} components would then have a spot
coverage of $\sim$70\% (as shown by our analysis of the primary in
Paper I), which is an extremely large fraction (effectively making
the stellar surface appear to be a very cool one covered with
hot spots, instead of the reverse).  While one may entertain such
extensive spottedness on one object, it strains belief to consider
it on both.}.  In this case, given that we find the same discrepancy,
both qualitatively and quantitatively, in both components, spots cannot
be responsible for preferentially depressing the \teff\ of the primary.

\noindent {\it (2)}\,  Opacity/EOS errors cause most of the discrepancy
instead.  If so, then the spot coverage on the primary would be far
too small to produce the \teff\ reversal: as noted in \S\ref{goals}, it is only
by ascribing the {\it entire} discrepancy in the primary to spots,
without considering any model errors, that we get the very large
coverage required by CGB07's spot theory.

\noindent {\it (3)}\,  Spots and model errors contribute equally
to the discrepancy in each component.  In this case, {\it both} of
the above conclusions would hold: the spot coverage on the primary
would be too small, and there would anyway be no marked difference
in spottedness between the components, again excluding spots as the
cause of the \teff\ reversal.

\noindent {\it (4)}\,  Spots cause the discrepancy in the primary,
but model errors, or some other effect, cause it in the secondary.
This is unlikely in the extreme, requiring a monumental coincidence.
In particular, the empirical $\log g$ of 2M0535A and B are nearly
identical, and their \teff\ are very similar as well (as evinced
by the empirical \teff\ ratio of $\sim$1.05, very close to unity).
Model errors would then have to be very prominent at the secondary's
\teff\ and $\log g$, but disappear over the small parameter jump to
the primary, which is improbable; simultaneously, the effects of such
errors on the secondary spectrum would have to coincidentally mimic
exactly the spot effects on the primary, which is even more unlikely.
The same argument applies to any other effect invoked for the secondary
but not the primary.

Together, these lines of reasoning imply that, while spots may
be present on both 2M0535A and B (and almost certainly are to some
extent, as shown by the lightcurve analysis of G09),
they {\it cannot} be responsible for {\it preferentially}
lowering the \teff\ of the primary by a large amount, and thus causing
the \teff\ reversal.

\subsection{General Implications for Magnetic Fields Causing \teff\ Suppression}

In light of this result, one might postulate that magnetic fields are not responsible for the temperature reversal at all, and perhaps heating due to tidal interactions (Heller et al.\,2010; Gomez Maqueo Chew et al.\,2012) is to blame instead.  However, with an orbital period of $\sim$10 days, and rotation periods of $\sim$3 days and $\sim$14 days in the primary and secondary respectively (Gomez Maqueo Chew et al. 2009), the two brown dwarfs in 2M0535-05 are sufficiently well separated and sufficiently non-synchronous that significant tidal interactions can be reasonably ruled out (Heller et al.\,2010, though a more thorough treatment of the coupled evolution of tidal effects and substellar structure is required to confirm this, as the latter authors note).

On the other hand, even if spots, caused by a local suppression of convection by magnetic
fields, are ruled out, fields may still produce the \teff\ reversal
by globally inhibiting convection in the interior of the primary.
Both CGB07 and MM09 have proposed such a mechanism.  The two theories
differ significantly in the interior field strengths invoked, however.  

MM09 apply a modified Gough-Taylor instability criterion, in which the magnetic energy basically scales as fraction ($\sim$1--10\%) of the total thermal energy, to derive required field strengths of order 10--100 MG in the interior.  In more recent work, they derive much smaller fields, but still $\sim$1 MG (MacDonald \& Mullan 2012).  However, these fields are orders of magnitude greater than the few 10s of kG interior fields, in equipartition with the kinetic (turbulent and convective) energy, suggested by recent simulations \citep{browning08,browning10}\footnote{Mullan \& MacDonald (2012) suggest that the simulations by \citet{browning08} probe relatively small rotational angular velocities, compared to fast rotating M dwarfs, and thus the field strengths derived in the simulations may be linearly scaled up with rotation rate.  However, the stars in these simulations are in fact quite close to saturation, so it is not clear that a linear increase in field strength with rotation is applicable (M.\,Browning, in prep.).}.  Moreover, interior fields $\gtrsim$\,1\,MG may also be buoyantly unstable; more detailed simulations are needed to check if this can be avoided.

Conversely, CGB07 present qualitative physical arguments suggesting that interior fields of $\sim$10\,kG (consistent with equipartition with the kinetic energy, and with the recent simulational results cited above) can seriously impede global convection in young brown dwarfs.  Within the context of mixing length theory, they find that a convective length scale parameter of $\alpha \ll 0.5$ is required to explain the \teff\ reversal in 2M0535AB.  Testing this, however, requires detailed full 3D simulations of cooling flows along flux tubes in a magnetized convecting medium, which is a considerable undertaking.  

Nevertheless, such simulations are essential to discriminate between the theories above, and test whether magnetic fields can indeed inhibit interior convection sufficiently to explain the observed \teff\ reversal.  The rapid pace of advance in simulational complexity \citep[e.g.][]{browning08,browning10} gives one hope that this will be become a reality in the not-too-distant future.  Concurrently, more observations are required to test the universality of such temperature suppression in young very low mass stars and brown dwarfs, not just in eclipsing systems but also in isolated objects.  The techniques used by \citet{morales08} for field low-mass stars offer a way forward here, though uncertainties in age-dependent luminosities for young objects will simultaneously present a problem in applying these methods.  However difficult a task, it must be tackled in order to really understand how magnetic fields affect the fundamental parameters of stars and brown dwarfs.

\acknowledgments  
S.M.\ acknowledges funding support from STFC grant ST/H00307X/1, and is indebted to the {\it International Summer Institute for Modeling in Astrophysics} (ISIMA) for providing the time and research environment to pursue this project. This work is supported in part by NSF grants AST-0607773 and AST-1009810 to K.G.S.

%\clearpage

\end{document}